\documentclass[11pt]{article}

\usepackage{hyperref,algorithmic,algorithm}
\usepackage{amsfonts}
\usepackage{graphicx,subfig}
\usepackage{amsmath,amsthm}
\usepackage{wrapfig}
\usepackage{amssymb}

\usepackage{epsfig}
\usepackage{amsmath}
\usepackage{amssymb}
\usepackage{psfrag}
\usepackage[absolute]{textpos}
\usepackage{setspace}

\newtheorem{thm}{Theorem}[section]
\newtheorem{lem}{Lemma} [section]
\newtheorem{define}{Definition}

\def\x{{\bf x}}

\newcommand{\beq}{\begin{equation}}
\newcommand{\eeq}{\end{equation}}
\newcommand{\bea}{\begin{eqnarray}}
\newcommand{\eea}{\end{eqnarray}}

\newcommand{\Prob}{\ensuremath{\mathbb{P}}}

\long\def\symbolfootnote[#1]#2{\begingroup%
\def\thefootnote{\fnsymbol{footnote}}\footnote[#1]{#2}\endgroup}

\def\x{{\bf x}}

\newcommand{\karis}{{(\eps,d,r_0,n)}}

\newcommand{\tr}{\text{rank}}

\newcommand{\la}{\lambda}
\newcommand{\A}{\mathcal{A}}
\newcommand{\B}{\mathcal{B}}
\newcommand{\N}{\mathcal{N}}
\newcommand{\PS}{\mathbb{S}}
\newcommand{\R}{\mathbb{R}}
\newcommand{\GG}{\mathcal{G}}

\newcommand{\E}{\mathbb{E}}
\newcommand{\X} {{\bf{X}}}
\newcommand{\s}{\star}
\newcommand{\eps}{\epsilon}
\newcommand{\sig}{\sigma}

\newcommand{\ts}{\text{Span}}

\begin{document}
\title{Subspace Expanders and Matrix Rank Minimization}

\author{ Amin Khajehnejad, Samet Oymak  and Babak Hassibi \\ California Institute of Technology, Pasadena CA 91125
\thanks{This work was supported in part by the National Science Foundation under grants CCF-0729203, CNS-0932428 and CCF-1018927, by the Office of Naval Research under the MURI grant N00014-08-1-0747, and by Caltech's Lee Center for Advanced Networking.}
}

\maketitle

\begin{abstract}
Matrix rank minimization (RM) problems recently gained extensive attention due to numerous applications in machine learning, system identification and graphical models. In RM problem, one aims to find the matrix with the lowest rank that satisfies a set of linear constraints. The existing algorithms include nuclear norm minimization (NNM) and singular value thresholding. Thus far, most of the attention has been on i.i.d. Gaussian measurement operators. In this work, we introduce a new class of measurement operators, and a novel recovery algorithm, which is notably faster than NNM. The proposed operators are based on what we refer to as subspace expanders, which are inspired by the well known expander graphs based measurement matrices in compressed sensing. We show that given an $n\times n$ PSD matrix of rank $r$, it can be uniquely recovered from a minimal sampling of $O(nr)$ measurements using the proposed structures, and the recovery algorithm can be cast as matrix inversion after a few initial processing steps.
\end{abstract}

%

\section{Introduction}
In the rank minimization problem, one would like to find a matrix $X$ with the lowest rank that satisfies a set of linear constraints $\A(X)$, often of smaller size than the number of matrix entries. This problem in its full generality is NP-hard. However, a number of recent papers have demonstrated that under certain circumstances, and when the unknown matrix is sufficiently low rank, RM can be solved via convex optimization programs, mainly nuclear norm minimization~\cite{Fazel,recht,recht_Weiyu,oymak}. Random linear Gaussian measurements are considered as a standard choice, for which certain recovery guarantees can be asymptotically proven for NNM. The two main conditions that certify the success of NNM are Restricted Isometry Property (RIP)~\cite{recht} and null space conditions~\cite{recht_Weiyu}. Based on the analysis initially developed by Stojnic for sparse recovery, tight analytical thresholds for NNM have been recently found by Oymak et al. and Chandrasekaran et al. using a ``escape through the mesh'' analysis of the null space conditions when measurement operators are i.i.d. Gaussian~\cite{oymak,venkat,mihailo}. In parallel, there have also been promising results on matrix completion problem as well, whereby one observes a subset of the entries of a low rank matrix, rather than linear combinations \cite{candes_MC,candes_Tao_MC}.

The RM problem is often regarded as a generalization of compressed sensing(CS), wherein the aim is to recover a sparse vector from a set of ill-posed linear measurements, represented by a wide measurement matrix $A^{m\times n}$ with $m<n$~\cite{rice}. It is now well understood that although certain random ensembles of measurement matrices (e.g. Gaussian, partial Fourier, etc.)  are legitimate choices for CS, carefully designed matrices can lead to additional benefits for the sparse compression/recovery. A few examples are faster encoding time and recovery algorithms for sparse matrices and in particular expander graphs~\cite{weiyu2,indyk,indyk_smp,amin}, higher recovery thresholds for  measurement matrices  based on Reed-Solomon codes \cite{Farzad,tarokh} etc. To the best of our knowledge, this point has not been studied in RM problem, where only random measurement ensembles (mostly Gaussian) have been studied. Do alternative recovery algorithms for RM problem exist, with  success guarantees on certain classes of carefully designed linear measurement operators?

%

In this paper, we introduce a new class of measurement operators, along with a novel recovery algorithm that is provably successful for the proposed operators, and is faster than NNM. The premise of the recovery algorithm relies on  the measurement operator $\A(\cdot):\mathbb{R}^{n\times n}\rightarrow \mathbb{R}^{m\times m}$ having three key properties: Hermitian, low density and rank expansion. Low density property means that $\A(X)$ can be described as the linear combination of $d$ linear operators $\A(X) = \sum_{i=1}^{d}\A_i(X)$, where $d$ is a constant, and for each $1\leq i\leq d$, the rank of $\A_i(X)$ is not larger than the rank of $X$. In fact, the low density operators that we introduce are characterized by only $O(mn)$ variables, as opposed to full i.i.d Gaussian linear measurements that require $(mn)^2$ variables. The interpretation of expansion is that $\A(\cdot)$ maps a sufficiently low dimensional subspace to a higher dimensional subspace, which is equivalent to mapping every positive semidefinite matrix $X$ to a positive semidefinite  matrix $\A(X)$ with rank greater that $c\cdot \tr(X)$, where $c>1$ is a constant. The contributions of this paper are thus threefold. We first prove that low density rank expander operators exist. We then provide a uniqueness result, in the sense that a sufficiently low rank PSD matrix $X$ is the unique PSD solution of the equations given by a high quality rank expander. We further propose a new recovery algorithm and provide theoretical recovery guarantees when the suggested rank expander measurement operators are exploited. Although most of the results are this paper are stressed for the cased of PSD matrices, we briefly mention a generalization of our results to the case of Hermitian matrices. This includes both the existence of high quality expanders operating on Hermitian matrices, and the solidity of the recovery algorithm. The validity of the proposed algorithm is verified by numerical simulations.
%
%
%
\section{Basic Definitions and Lemmas}
Let $\PS^n$ denote the space of Hermitian  matrices of size $n\times n$, and $\PS^n_+$ denote the set of positive semidefinite (PSD)  matrices. An orthogonal projection is a matrix $P\in\PS^n_+$ with $P^2=P$. We say that $U\in\R^{n_1\times n_2}$ is a partial unitary matrix if $U^TU=I$, i.e. the columns of $U$ form an orthonormal set. Notice that $UU^T$ is an orthogonal projection. Let $\eta_+(X),\eta_-(X),\eta_0(X)$ denote the number of positive, negative and zero eigenvalues of $X$, respectively. Also for a Hermitian matrix $X$, let $X_-$ and $X_+$ denote the PSD matrices induced by the negative and positive eigenvalues of $X$ respectively, i.e. $X=X_+-X_-$.


For a given matrix $X\in\R^{n_1\times n_2}$, $\lambda_i(X)$ and $\sig_i(X)$ denotes $i$'th largest eigenvalue and the $i$'th  largest singular value, respectively. The nuclear norm, spectral norm and Frobenius norm operators are denoted by $\|\cdot\|_\s$,  $\|\cdot\|$ and  $\|\cdot\|_F$, respectively, and are defined by $\|X\|_\s=\sum_{i=1}^{\min\{n_1,n_2\}}\sig_i(X)$, $\|X\|=\sig_1(X)$ and $\|X\|_\s=\left(\sum_{i=1}^{\min\{n_1,n_2\}}\sig_i^2(X)\right)^{1/2}$. In addition, we define $\ts_C(X),\ts_R(X)$ to be the linear spaces spanned by the columns and rows of $X$, respectively.

A function $f:\R^n\rightarrow \R$ is called $L$-Lipschitz if $|f(x)-f(y)|\leq L\|x-y\|_{\ell_2}$, for every $x,y$.  For hermitian matrices $A,B$, $A\succeq B$ means that $A-B$ is positive semidefinite. For a linear operator $\A(\cdot)$ acting on a linear space, we denote the null space of $\A$ by $\N(\A)$, i.e. $W\in\N(\A)$ iff $\A(W)=0$. We denote by $\GG(d_1,d_2)$ the ensemble of real $d_1\times d_2$ matrices in which the entries are i.i.d. $\N(0,1)$ (zero-mean, unit variance Gaussian).

The following lemmas are crucial to the technical discussions of this paper. Their proofs are skipped due to space considerations.


\begin{lem}
\label{lem:lipsc}
Let $f(X)$ be a function on matrices in the following form: $f(X)=\sum_{i=1}^m a_i \sig_i(X)$ for some real constants $\{a_i\}_{i=1}^m$. Then $f(X)$ is a $\sqrt{\sum_{i=1}^m a_i^2}$ Lipschitz function of $X$.
\end{lem}

\begin{lem}
\label{lem:gauslip}
{\bf{(A Gaussian concentration inequality, \cite{talagrand})}}\\
Let $\x$ be drawn from $\GG(n,1)$ and $f : \R^n \rightarrow \R$ be a function with Lipschitz constant $L$. Then, we have the following concentration inequality
\vspace*{-2pt}
\begin{equation}
\Prob(|f(x)-\E f(x)|\geq t)\leq 2\exp(-\frac{t^2}{2L^2})
\end{equation}
\end{lem}

\begin{lem}
\label{lem:weyl}
{\bf{(Weyl's Inequalities, \cite{Bhatia})}}\\
Let $A,B\in\PS^n$. Then:
\begin{align}
&\la_j(A+B)\leq\la_i(A)+\la_{j-i+1}(B)~~~\forall~i\leq j\\
&\la_j(A+B)\geq\la_i(A)+\la_{j-i+n}(B)~~~\forall~j\leq i
\end{align}
\end{lem}


\section{Rank Expanders and Proposed Operators}

\begin{define}
\label{def:expander}
Let $\A:\R^{n\times n}\rightarrow \R^{m\times m}$ be a linear operator with $m<n$. For $d >0, 0\leq \eps<1, 1 \leq r_0\leq n$, we say that $\A$ is an unbalanced $\karis$-rank expander, if it satisfies the following conditions:
\begin{enumerate}
\item For every $X\in\PS^n_+$,  $\A(X)\in\PS_+^m$
\item For every $X\in\PS^n$, $\tr(\A(X)) \leq d\cdot \tr(X)$
\item For all orthogonal projections $P$ with $\tr(P)=r\leq r_0$, $rd\geq \tr(\A(P)) > (1-\eps)rd$          
\end{enumerate}
\end{define}
This definition is inspired by the definition of unbalanced expander graphs, that maintain similar properties with respect to positive vectors (instead of PSD matrices) and with $\ell_0$-norm (instead of rank). An unbalanced $d$-regular $(k,\epsilon)$-expander graph is a bipartite graph with $n$ nodes on the left and $m$ nodes on the right, and regular degree $d$ for left hand side nodes, such that every subset $S$ of left nodes with $|S|\leq r_0$ has a neighborhood $N(S)$ of size at least $|N(S)|\geq (1-\epsilon)|S|d$. Unbalanced expander graphs have been proven to have elegant properties that make them suitable for sparse vector recovery (a.k.a. compressed sensing) in addition to being useful as parity check matrices for error correcting codes. With that in mind, one might be inspired to generalize the notion of expander graphs to subspace (rank) expanders, in order to obtain operators that can be used in low rank matrix recovery. The following lemma is immediate.
\begin{lem}
\label{lem:addition1}
If $\A(.)$ is an $\karis$-rank expander then, for every $X\in\PS^n_+$ with rank $r\leq r_0$, we have $rd\geq\tr(\A(X))\geq (1-\eps)rd$
\end{lem}
We now move on to describe the proposed measurement structures. Afterwards, we prove that these constructions indeed result in rank expanders and that the expansion property allows us to find alternative fast reconstruction algorithms for the RM problem.
\subsection{Proposed Measurement Operator}
Let $G_1,\dots G_d\in\R^{m\times n}$ be matrices to be specified later. The proposed measurement operator $\A(\cdot)$ has the following low density form:
\vspace*{-3pt}
\beq
\label{eq:proposed}
\A(X)=\sum_{i=1}^d G_iXG_i^T
\eeq
\noindent Where $X\in R^{n\times n}$. We will prove that upon appropriate choices of $G_i$'s, $\A(\cdot)$ is an unbalanced rank expander. It is easy to check that with this choice of $\A(.)$, conditions 1 and 2 of  Definition \ref{def:expander} are  immediately satisfied.
Furthermore, as long as the $G_iXG_i^T$'s are almost incoherent, one would expect their ranks to add up. In particular, it is easy to show that when $X$ is fixed and $\{G_i\}_{i=1}^d$ are drawn i.i.d. from $\GG(m,n)$, we have
\beq
\label{full_prob}
\Prob(\tr(\A(X))=\min\{d\times\tr(X),m\}) = 1.
\eeq
\noindent However, the challenge of condition 3 is in the fact that the rank expansion property must hold for every sufficiently low rank $X$. Let $X\in\PS^n_+$ and $X^{1/2}$ denote an arbitrary square root of $X$ (i.e. $X=X^{1/2}X^{T/2}$). Note that $\A(X)$ can be written in the following form:
\vspace*{-3pt}
\beq
\A(X)=\left(G_1X^{1/2}~\dots~G_dX^{1/2}\right)\left(G_1X^{1/2}~\dots~G_dX^{1/2}\right)^T
\eeq
\vspace*{-3pt}
It then follows that
\vspace*{-2pt}
\beq
\label{eq:ease}
\tr(\A(X))=\tr\left(G_1X^{1/2}~G_2X^{1/2}~\dots~G_dX^{1/2}\right)
\eeq
\noindent For analyzing the rank expansion property, we can thus limit ourselves to the form (\ref{eq:ease}).

\subsection{Existence of Rank Expanders}
Our goal is to prove the existence of high quality (small $\eps$) rank expanders for certain regimes of $d,\eps,r_0$ and $n$. Based on Lemma \ref{lem:addition1}, we can restrict our attention to $X$ being an orthogonal projection of rank at most $r_0$. Our analysis is for the case when $G_i$'s are chosen i.i.d. from $\GG(m,n)$ \footnote{Existence of expanders using other ensembles of matrices $G$, and in particular sparse matrices, shall remain as an interesting open problem}. The main existence theorem is the following:

\begin{thm} {\bf{Existence of Rank Expander}}\label{mainthm}
For any $0<\eps<1$ there are constants $C_1$ and $C_2$ so that for any $n$ and $r_0\leq n$, whenever  $m=\sqrt{C_1C_2 nr_0}$ and $d=\sqrt{\frac{C_2 n}{C_1r_0}}$ and $\{G_i\}_{i=1}^d$'s are independent instances of $\GG(m,n)$, the operator $\A(X)=\sum_{i=1}^dG_i XG_i^T$ is an $\karis$ expander with probability at least $1-\exp(-\Omega(n))$.
\end{thm}

Before explaining the technicalities involved in the proof of the above theorem, consider the following argument.   Given $\A:\R^{n\times n}\rightarrow \R^{m\times m}$, suppose for all $X\in\PS^n_+$ with $\tr(X)\leq r^*$, $X$ can be uniquely decoded from $\A(X)$, say by exhaustive search. It then follows that $\A$ has to be injective on the restricted domain  $\{X\in\PS^n_+:\tr(X)\leq r^*\}$.
%
It will soon become apparent in the sequel that given an $\karis$ expander, for $r^*= r_0/2$ this condition holds. A simple argument counting the degrees of freedom of the low rank domain and the corresponding range of $A$  reveals that the problem parameters should satisfy the following relationship.
\beq
\label{simpineq}
m =\Omega(\sqrt{nr_0}),~md = \Omega(n),~dr_0=O(m)
\eeq
\noindent In fact, it turns out that Theorem \ref{mainthm} is true as long as (\ref{simpineq}) holds asymptotically, which implies the optimality of the number of measurements in the suggested expander operators. For the proof, we set $m=C_1dr_0$ and $dm=C_2n$ where $C_1>1$, $C_2>1$ will be the constants in Theorem \ref{mainthm}.

%

\begin{proof}[Proof sketch of  Theorem \ref{mainthm}.] The proof is based on three major technical steps.\\
\noindent {{\bf{Step 1:}}} We consider an $\eps_0$ cover with operator norm $\|\cdot\|$ over the set of orthogonal projections of rank $r\leq r_0$. From \cite{szarek}, we know that there is such a cover of size at most $M=\left({C_0}/{\eps_0}\right)^{nr}$, which we denote by $\{U_iU_i^T\}_{i=1}^M\in\PS^n_+$, with $U_i\in\R^{n\times r}$. Also, we first focus on a particular rank $r\leq r_0$, and later union bound the undesirable probability over all values of  $r\leq r_0$.

\noindent {{\bf{Step 2:}}} Now consider a $U_i$ from the $\epsilon_0$ cover. Denote $\B(U_i)=[G_1 U_i~\dots~G_dU_i]\in\R^{m\times dr}$. Since $\tr(\A(U_iU_i^T))=\tr(\B(U_i))$,  we can focus on $\B(U_i)$. Note that since $U_i$ is a fixed partial unitary, due to the unitary invariance of i.i.d. Gaussian matrices, $\B(U_i)$ has i.i.d. Gaussian distribution. Now define the function $f(X)=\sum_{i=(1-\eps)dr+1}^{dr} \sig_i(X)$, for $X\in\R^{m\times rd}$. Notice that $f(X)>0$ implies $\tr(f(X))>(1-\eps)dr$, because it means some of the smallest $\eps dr$ singular values of $X$ are nonzero. On the other hand $f(X)$ is a linear function of the singular values of $X$, and thus it satisfies the Lipschitz condition of Lemma \ref{lem:lipsc}. Since $\B(U_i)$ is Gaussian, we can apply Lemma \ref{lem:gauslip} to get the following concentration bound
\vspace*{-5pt}
\beq
\label{concent1}
\Prob(f(\B(U_i))<\delta \frac{e}{\sqrt{m}})<\exp(-\frac{(C_3-\delta)^2 e}{2})
\eeq
where $C_3=1-\sqrt{1/C_1}$ and $e=\eps drm$. Here $\delta$ serves as a safety margin, in order to account for the perturbation $P-U_iU_i^T$, when we consider a certain orthogonal projection $P$, which we know is the proximity of some  $U_iU_i^T$ of the cover, i.e. $\|P-U_iU_i^T\|<\eps_0$. In other words, as will be shown in the next step, lower bounding $f(\B(U_i))$ certifies that $\tr(\A(P))$ will also be large, and thus (\ref{concent1}) is basically determines a  lower bound on the probability of failure. The exponent of the righthand side of (\ref{concent1}) is $e=\eps drm=\eps C_2 rn$ which is proportional to the exponent of the size of the cover $\log(M)=O(rn)$. Consequently, with careful choices of parameters, using a union bound on failures, we can make sure $f(\B(U))\geq\delta \frac{e}{\sqrt{m}}$ for all $U\in\{U_i\}$ w.h.p. In particular we need:
\vspace*{-4pt}
\beq
\label{cond1}
(C_3-\delta)^2 \eps C_2>(\log C_0-\log \eps_0)
\eeq

\noindent {{\bf{Step 3:}}} Now it remains to show that if $E=P-P_i$ is a perturbation on $P_i$ which makes $\tr(P)\leq (1-\eps)dr$,
 then $\|E\|$ has to be large because $f(\B(U_i))$ is large. In particular, showing $\|E\|>\eps_0$ will finish the proof since we know that $\|P-P_i\|\leq\eps_0$. In order to show this step, we make use of Lemma \ref{lem:weyl} to find $\lambda_{i+(1-\eps)dr}(\A(P_i))\leq \lambda_i(\A(E))$ and hence to deduce that $\|\B(E_+^{1/2})\|_\s\geq f(\B(U_i))$. We carry out some more arguments to upper bound $\|\B({E_+^{1/2}})\|_\s$ in terms of $\|E\|$ to get a contradiction as long as $\delta\sqrt{\eps}>2\sqrt{\eps_0}~(*)$ holds. Finally, we conclude that whenever $(*)$ and the condition (\ref{cond1}) are satisfied, with high probability $f(\B(\sqrt{P}))>0$ for all projections $P$ with rank $r$, which implies $\tr(\A(P))> (1-\eps)dr$ as desired. In particular, sufficiently large values of $C_1,C_2$ will do the job. Also, inequalities (\ref{simpineq}) will similarly work since increasing $C_1,C_2$ only improve the conditions.
\end{proof}

\vspace*{-3pt}
\section{Fast Recovery  Algorithm}
Before presenting the main algorithm, we will provide some results about low rank positive semidefinite matrix recovery. Specifically, we emphasize on a uniqueness result concerning PDS matrices and the rank expander operators. Suppose a matrix $X_0\in\PS^n_+$ and a linear operator $\A(\cdot)$ are given, and we ask under what conditions $X_0$ is the unique (PSD) inverse image of  $\A(X_0)$, i.e. it is possible to recover $X_0$ by simply characterizing the set $\{X~|~X\succeq 0, \A(X)=\A(\X_0)\}$. The following lemma which is adopted from \cite{weiyu,oymak} provides an answer to this question.
\vspace{-1pt}
\begin{lem}
\label{lem:psdstr}
Any PSD matrix $X$ of rank at most $r$ is the unique PSD inverse image of $\A(\X)$, if and only if every  nonzero Hermitian $W\in\N(\A)$ has at least $r+1$ negative eigenvalues.
\end{lem}
Now, we  explain how rank expanders can facilitate the existence of the condition in Lemma \ref{lem:psdstr}.
\begin{lem}
\label{lem:expstr}
Let $\A(.)$ be an $\karis$-rank expander with $\eps<1/2$. Then for every nonzero Hermitian $W\in\N(\A)$ we have that $\eta_-(W)> r_0/2$.
\end{lem}
\begin{proof}
Any $W\in\N(\A)$, write $W=W_+-W_-$. Since $\A(W)=0$ we have $B=\A(W_+)=\A(W_-)$. Assume $\eta_-(W)\leq r_0/2$. Let $r=\min\{\eta_-(W),\eta_+(W)\}$. Then $dr\geq \tr(B)$. On the other hand $\tr(W_++W_-)=\tr(W_+)+\tr(W_-)\geq 2r$ hence we have $\tr(\A(W_++W_-))\geq 2(1-\eps)dr$ since $2r\leq r_0$. Note that $\A(W_++W_-)=\A(W_+)+\A(W_-)=2B$. It follows that $dr\geq \tr(B)=\tr(2B) \geq 2(1-\eps)dr\implies 1\geq 2(1-\eps)\iff \eps\geq 1/2$, which is a contradiction.
\end{proof}

The combination of Lemmas \ref{lem:psdstr} and \ref{lem:expstr} suggests that by using an $\karis$-rank expander with $\eps<1/2$ as a measurement operator, one can guarantee that every PSD matrix $X_0$ of rank at most $r_0/2$ is the unique PSD solution to the measurements $\A(X_0)$. Therefore, every program (e.g. SDP) that can identify a point in the feasible set  $\{X~|~X\succeq 0, \A(X)=\A(\X_0)\}$ successfully returns $X_0$. Quite Interestingly, for the case of high quality expanders (small $\eps$), we propose an alternative method for identifying one feasible point,  which by the token of the aforementioned uniqueness argument, can successfully recover low rank PSD matrices. The algorithm is based on using the expansion property of the suggested linear operator, and the fact that the original matrix is low rank. The key point is that for rank expanders, the original under-determined system can be equivalently transformed to an over-determined linear system, after a few simple processing steps, that mostly involves taking SVD and finding null spaces. The new system of linear equation can then be solved by matrix inversion. This routine is described in Algorithm \ref{alg:1}, and is in fact the generalization of a positive sparse vector recovery algorithm elaborated in \cite{amin}. The next theorem provides a guarantee for the success of the proposed algorithm.
\begin{algorithm}[t]
\caption{\small{Reconstruct a low rank PSD matrix $X$ from under-determined linear measurements
$Y=\sum_{i=1}^d A_iXA_i^*$.}}
\begin{algorithmic}[1]
\STATE \textbf{Input:}
\STATE Constant integer $d \geq 1$.
\STATE Matrices $A_i\in \mathbb{R}^{m\times n}, 1\leq i \leq d$, and $Y\in \mathbb{R}^{m\times m}$.
\STATE \textbf{Output:}
\STATE  Low rank PSD matrix $X$.
\STATE \textbf{Initialize}
\STATE Compute $Y = S\Sigma S^*$, with $S$ full column rank (SVD).
\STATE Set $P=I-SS^*$.
\STATE Set $Q:=\text{Null}\big((PA_1)^T,\dots,(PA_d)^T\big)^T$.
\STATE Compute $B_i = A_iQ$, and set $M = \sum_{i=1}^d B_i\otimes B_i$.
\STATE Find $X\in \mathbb{R}^{n\times n}$ with $\text{vec}(X) = (Q\otimes Q)M^{\dag}vec(Y)$.
\end{algorithmic}
\label{alg:1}
\end{algorithm}

\vspace*{-1pt}
\begin{thm}{{\bf{PSD Recovery.}}}
\label{thm:recovery_guarantee}
If the operator $\A(X) = \sum_{i=1}^d A_iXA_i^T$ is a $\karis$-rank expander with $\epsilon < 1/2$, then for every $k\leq r_0(1-\eps)$, every PSD matrix $X$ of rank $k$ can be perfectly recovered from $\A(X)$ using Algorithm \ref{alg:1}.
\end{thm}
In order to prove Theorem \ref{thm:recovery_guarantee}, we first prove the following lemma.
\vspace*{-1pt}
\begin{lem}
\label{lem:zero_ident}
Suppose that the operator $\A(X) = \sum_{i=1}^d A_iXA_i^T$ is an unbalanced $\karis$-rank expander with $k/(1-\eps)<m<n$, where $A_i$'s are $m\times n$. Also suppose that $X\in \PS^n$ has rank $k \leq r_0(1-\eps)/2$. Further, let $\mathcal{S}$ be the linear space of all $n\times 1$ vectors $u$ such that $\ts_C\{\A(uu^T)\}\in\ts_C\{\A(X)\}$. Then $\text{dim}(\mathcal{S})< k/(1-\eps)$.
\end{lem}
\begin{proof}
If $\text{dim}(\mathcal{S})\geq  k/(1-\eps)$, then we can find an orthonormal matrix $U$ of size $n\times k/(1-\eps)$ such that all of its columns are in $\mathcal{S}$. Therefore, by definition $\ts_C\{\A(UU^T)\}\in\ts_C\{\A(X)\}$, and thus $\text{rank}(\A(UU^T)) \leq \text{rank}(\A(X)) \leq k\cdot d$. However, since $k/(1-\eps) < r_0$, from the definition of rank-expander, we must have $\text{rank}(\A(UU^T)) > d(1-\eps)\text{rank}(UU^T) = k\cdot d$, which is a contradiction.
\end{proof}
\begin{proof}[Proof of Theorem \ref{thm:recovery_guarantee}]
Let $\mathcal{S}$ be as defined in Lemma \ref{lem:zero_ident} with $\text{dim}(\mathcal{S})=r$, and $Q^{n\times r}$ be a basis for $\mathcal{S}$, and let $X = X^{1/2}X^{T/2}$. It is easy to check that the columns of $X^{1/2}$ are all in $\mathcal{S}$, and thus $X^{1/2} = QV^{1/2}$, for some unknown $V^{1/2}$ of size $r\times k$. Therefore, we can write:
\vspace*{-5pt}
\beq
Y = \A(\X) = \sum_{i=1}^{d}A_iQV^{1/2}V^{T/2}Q^TA_i^T
\eeq
\vspace*{-5pt}
\noindent or equivalently
\vspace*{-5pt}
\beq
\label{eq:lin. eq.}
\text{vec}(Y) = \left(\sum_{i=1}^{d} B_i\otimes B_i^T \right)\text{vec}(V)
\eeq
\noindent where $B_i = A_iQ$. (\ref{eq:lin. eq.}) is a linear system of equations with $m^2$ equations and $r^2$ unknowns. Moreover, from Lemma \ref{lem:zero_ident}, we know that $r<k/(1-\eps) < m$. In addition, from Lemma \ref{lem:expstr},  the solution to the system of linear equations in (\ref{eq:lin. eq.}) is unique and must be $V=X$, since otherwise $X-V$ is in the null space of $\A(\cdot)$, and has at most $r<r_0/2$ negative eigenvalues. Therefore, (\ref{eq:lin. eq.}) is an over-determined linear system and can be solved by matrix inversion as given in line 11 of Algorithm \ref{alg:1}.
\end{proof}
\vspace*{-8pt}
\section{Extension to Hermitians}
In this section, we briefly extend the results of  the previous sections to the case of Hermitian matrices. Specifically, we prove the existence of  expanders for Hermitian matrices (rather than only PSD matrices, which was the case discussed previously), and then state a theorem certifying the success of Algorithm \ref{alg:1} for these classes of low rank matrices.
\vspace*{-5pt}
\subsection{Expansion}
\begin{lem} {{\bf{Expansion for Hermitians.}}} Assume $\A$ is a $\karis$ expander and let $X$ be Hermitian with $\tr(X)\leq r_0$. Then, $\tr(\A(X))\geq (1-4\eps)d\cdot\tr(X)$.
\label{lem:herm}
\end{lem}
The formal proof of the above lemma is skipped, but it is mostly based on some dimension counting arguments and the fact that $\tr(A+B)=\tr(A)+\tr(B)$ is equivalent to $\ts_C(A)\cap\ts_C(B)=\emptyset$ and $\ts_R(A)\cap\ts_R(B)=\emptyset$ (see e.g. \cite{callan}).
\vspace*{-2pt}
\subsection{Recovery}
\vspace*{-2pt}
Assume $X\in\PS$ with $\tr(X)=r\leq r_0$. With Gaussian measurements, $\tr(\A(X))=rd$ almost surely, due to (\ref{full_prob}) and the result of \cite{callan}. Notice that if the  eigenvalue decomposition of $X$ is $X=\sum_{i=1}^r \lambda_i u_iu_i^T$, then $\A(X)=\lambda_i\sum_{i=1}^r \A(u_iu_i^T)$, and using $\tr(\A(X))=r\times d$ and $\tr(\A(u_iu_i^T))\leq d$, it follows that $\ts_C(\A(u_iu_i^T))\subset\ts_C(\A(X))$ for all $i\leq r$. Consequently, similar to the case of PSD matrices, we need to find the space of $u$ such that $\ts_C(\A(uu^T))\subset\ts_C(\A(X))$. All technical steps follow identically and similar to the PDS case, and we can assert that sufficiently low rank Hermitian solutions to (\ref{eq:lin. eq.}) are unique, as follows.
\vspace*{-3pt}
\begin{thm} {{\bf{Hermitian Recovery.}}}
Let $X_0\in\PS$ with $\tr(X_0)\leq r_0(1-\eps)$. Suppose $\A(\cdot)$ is as described in Theorem \ref{mainthm}. With probability at least $1-\exp(-\Omega(n))$, $X_0$ can be perfectly recovered from $\A(X_0)$ by Algorithm \ref{alg:1}.
\end{thm}
\vspace*{-4pt}
Note that although the same algorithm works for both cases of PSD and Hermitian matrices, there is a significant difference. In Theorem \ref{thm:recovery_guarantee}, ``all'' $X\in\PS^n_+$ with sufficiently rank low are recoverable, whereas a similar fact is true for ``almost all'' Hermitian matrices of low rank. These notions are often distinguished in the literature  by the terms ``strong'' recovery and ``weak'' recovery, respectively. Furthermore, note that for the case of PSD recovery, one can alternatively use convex optimization to find the unique PSD solution. However for the recovery of Hermitians, we do not know of any other method but our proposed Algorithm \ref{alg:1}.
\vspace*{-4pt}
\section{Simulation Results}
\begin{figure}[h]
\centering
  \includegraphics[width= 0.9\textwidth]{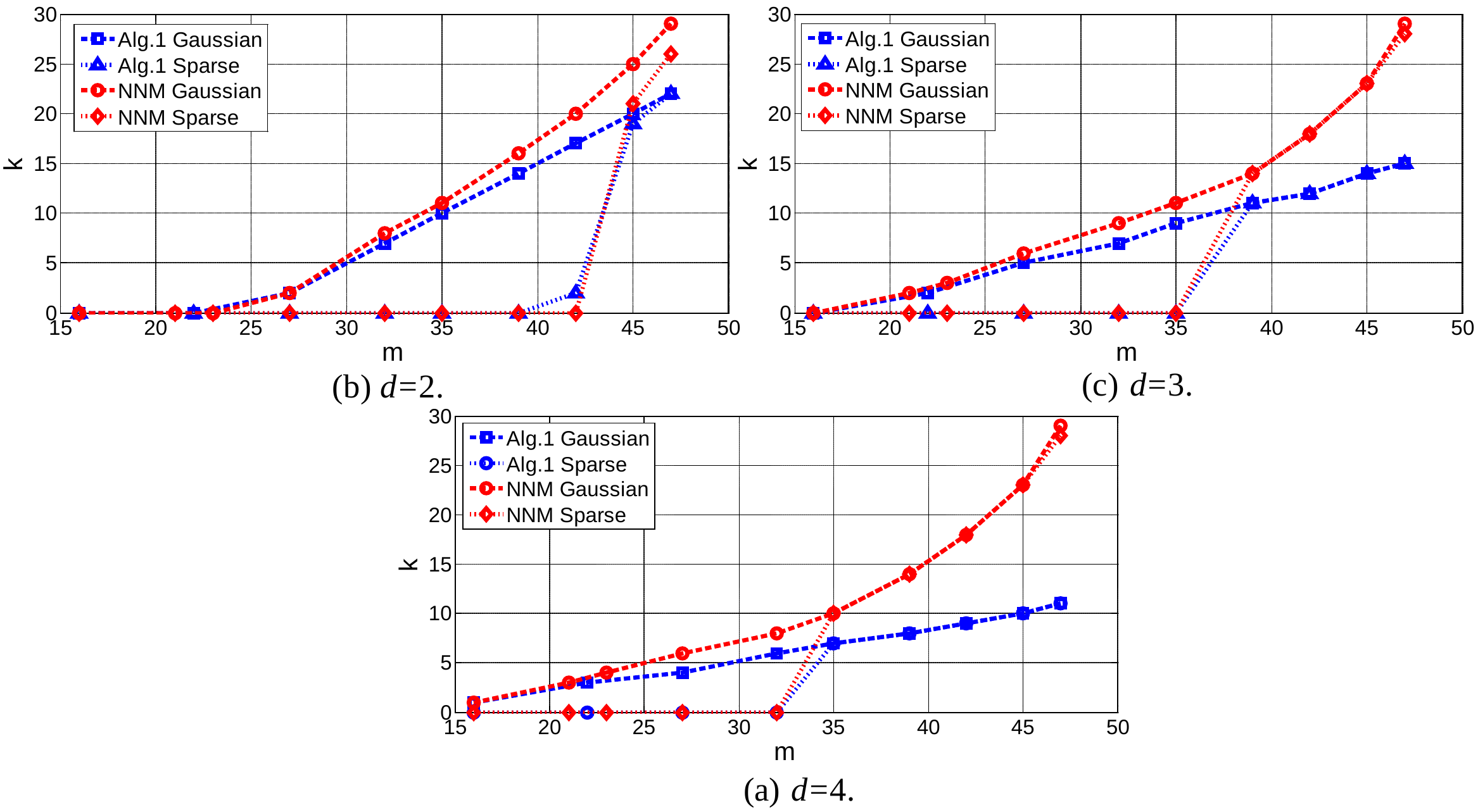}
 \caption{\scriptsize{Empirical recovery thresholds of Alg. \ref{alg:1} and NNM for $50\times 50$ matrices with linear operators. Measurements are $m\times m$ and $k$ is the rank.}}
  \label{fig:curves}
\end{figure}

\vspace*{-3pt}
Numerical simulations were performed to verify the validity of Alg. \ref{alg:1}. We used $n=50$, and linear   measurement operators in the form of (\ref{eq:proposed}) with $d=2,3,4$, and two types of distributions for $G_i$'s: 1)  i.i.d Gaussian matrices, and 2) sparse matrices where every row of each $G_i$ has exactly one 1 in a random location. We did not explicitly prove that the sparse constructions expanders. However, such low density structures are of high practical interests. The resulting curves of successful recovery thresholds are given in Figure \ref{fig:curves}. In all of our simulations for Gaussian matrices, the transition between successful recvery and failure was very sharp, i.e. either failed all times or succeeded, depending on the number of measurements and the rank of $X$.   Figure \ref{fig:curves} illustrates the empirical transition phase or the recovery threshold of Alg. \ref{alg:1}. The corresponding curves for sparse matrices are also shown in Figure \ref{fig:curves}. The transition for sparse matrices was  not as sharp as Gaussians.  On the same curves, the performance of the standard trace minimization with nonnegativity (PSD) constraint is also displayed.  Observe that the performance of the proposed algorithm is very comparable to the convex relaxation method. In addition, the curves for sparse measurements collapse into the same curves as the dense measurements for sufficiently large $m$. In practice, Alg. \ref{alg:1} is extremely faster than NNM. To give an example, for simulations in MATLAB on a 2.4 GHz Intel Core i5 with 4 GB RAM, and for the case of $d=2,m=39,k=10$ with Gaussian $G_i$'s, each reconstruction took on average 34s for the NNN, while it was only about 0.05s for Alg. \ref{alg:1}. For sparse structures, it took about 1.8s for NNM and 0.05s for Alg.\ref{alg:1}. NNM was solved via the SEDUMI toolbox.
\vspace*{-5pt}


\end{document}